\documentclass{jpconf}

\usepackage{graphicx}

\newcommand{\Cee}{\mathcal{C}}

\renewcommand{\mat}[1]{\mathsf{#1}}
\renewcommand{\vec}[1]{\mathbf{#1}}
\renewcommand{\Re}{\mathrm{Re}}

\begin{document}

\title{An introduction to phase transitions in stochastic dynamical
systems}

\author{R A Blythe}

\address{School of Physics, University of Edinburgh, Mayfield Road,
  Edinburgh EH9 3JZ}

\ead{R.A.Blythe@ed.ac.uk}

\begin{abstract}
We give an introduction to phase transitions in the steady states of
systems that evolve stochastically with equilibrium and nonequilibrium
dynamics, the latter defined as those that do not possess a
time-reversal symmetry.  We try as much as possible to discuss both
cases within the same conceptual framework, focussing on dynamically
attractive `peaks' in state space.  A quantitative characterisation of
these peaks leads to expressions for the partition function and free
energy that extend from equilibrium steady states to their
nonequilibrium counterparts.  We show that for certain classes of
nonequilibrium systems that have been exactly solved, these
expressions provide precise predictions of their macroscopic phase
behaviour.
\end{abstract}

\section{Introduction}

It is highly desirable that we properly understand the role of
microscopic dynamics in shaping the macroscopic properties of
many-body interacting systems.  This was highlighted many times during
the recent Summer School on ``Ageing and the Glass Transition''
\cite{AGT05} where the microscopic origins of glassy behaviour were
much debated \cite{Kob,Kruger}.  Macroscopically, one sees at the
glass transition diverging relaxational timescales---manifestly
dynamical quantities---e.g., the viscosity of a fluid increasing by
many orders or magnitude \cite{DS01,Angell95}.  One prominent
macroscopic effect in the glassy phase is ageing \cite{BCKM97}, where
experimental observables depend on the overall time since the glass
was formed.  A microscopic picture often used to explain such
behaviour has particles being caged by their nearest neighbours, thus
impeding relaxation to equilibrium.  Meanwhile, spin glass systems
\cite{Young97,Vincent} exhibit exotic memory and rejuvenation effects
in their glassy phase.  Here, a microscopic dynamical explanation
involves the exploration of complex free-energy landscapes with many
nested minima.  More generally there is a drive to understand
complexity in all manner of dynamically interacting systems, such as
societies, ecosystems, geological structures and so on (see, e.g.,
\cite{BarYam97} for an introduction).

If one is dealing with a system that is not subject to a driving
force, equilibrium statistical mechanics provides a complete theory
for deriving macroscopic properties from a microscopic model.  This
success perhaps gives us hope that we may be able to extend such
theories to \emph{nonequilibrium} systems, e.g., the glassy and
complex systems described above and which are typically couched in
terms of microscopic dynamical rules.  A problem here is that the
equilibrium theory makes little reference to the latter, other than
the tendency for equilibrium dynamics to maximise disorder.  To
understand the consequences of enforcing \emph{nonequilibrium}
dynamics---specifically those that drive the system away from thermal
equilibrium---I focus exclusively on the macroscopic phenomenon of the
\emph{phase transition}.  In order to make connections between phase
transitions in equilibrium and nonequilibrium steady states, I shall
revisit the former from the viewpoint of a dynamical phase-space
exploration.  It is therefore necessary to be clear about the
distinction between equilibrium and nonequilibrium dynamics.  This is
where I begin.

\section{Equilibrium and nonequilibrium stochastic dynamics}
\label{stoch}

Consider a system of classical particles in thermal equilibrium with a
heat bath at inverse temperature $\beta = 1/kT$.  Although the
trajectories of these particles can in principle be predicted from
Newton's deterministic equations of motion, it is appropriate to use a
\emph{stochastic} description of the dynamics due to the complexity in
the dynamics resulting from the vast number of collisions that take
place (see, e.g., \cite{vanKampen92,Zaslavsky05}).  A further
indeterminacy lies in fact that the precise nature of the interaction
between the system and the heat bath is often unspecified.

We express these stochastic rules using the probabilities $M_{\delta
t}(\Cee\to\Cee',t)$ that the system evolves from a configuration
(point in phase-space) $\Cee$ at time $t$ to reach a new configuration
$\Cee'$ after a time interval $\delta t$.  Note that one can employ a
discrete-time stochastic model even when the underlying dynamics are
continuous.  We now identify four properties this stochastic process
should possess in order to model the evolution of a system at
equilibrium.
\begin{enumerate}
\item\label{Markov}\textit{Markov property} --- The transition
probabilities $M_{\delta t}(\Cee\to\Cee',t)$ are assumed not to depend
on the history of the process prior to time $t$.  This is appropriate
because, given the phase-space coordinates $\Cee$ at time $t$ one
could, in principle, calculate the transition probabilities without
this historical information.
\item\label{tti}\textit{Time translational invariance} --- When a
system is not being driven by an external force, the underlying
equations of motion are time-translationally invariant, and so we can
drop the dependence of the transition probabilities on time.
\item\label{ergo}\textit{Ergodicity} --- An important property of
many-body dynamical systems is that of ergodicity, in which a quantity
averaged over a single trajectory equals that taken over the
stationary distribution of configurations obtained from many different
initial conditions \cite{Zaslavsky05}.  In the context of stochastic
processes in a finite state space, ergodicity is the property of there
being a unique stationary distribution that is converged on from every
initial condition and for every configuration to be represented by a
nonzero probability in this distribution\footnote{Actually, one can
find in the literature a number of different definitions of
ergodicity.  The one given here is the most restrictive and agrees
with that found in, say, \cite{Kelly79}.}.
\item\label{trsprop}\textit{Time reversal symmetry} --- At
equilibrium, the ensemble of realisations of the dynamics running
forwards in time cannot be distinguished from that of those running
backwards.  This is because the underlying equations of motion have
this property \cite{vanKampen92}.  Furthermore, there is no ``arrow of
time'' once the maximum-entropy equilibrium state has been reached.
\end{enumerate}
There is only enough space here to highlight the most important
consequences of these properties; for full details, the reader
should consult textbooks on stochastic processes, such as
\cite{vanKampen92,Kelly79}.

Together, properties (\ref{Markov}) and (\ref{tti}) imply that we can
describe the evolution of the probability distribution over
state-space $P(\Cee)$ in terms of a \emph{master equation}
\begin{equation}
\label{master1}
P(\Cee, t+\delta t) = \sum_{\Cee'} P(\Cee', t) M_{\delta t}(\Cee' \to \Cee) \;.
\end{equation}
In order for a stochastic process to be ergodic [property
(\ref{ergo})], the dynamics must allow every configuration $\Cee'$ to
be reached from any other configuration $\Cee$ after a fixed (but
possibly large) number of transitions.  Whether ergodic or not,
probability must be conserved by the dynamics: $\sum_{\Cee'} M_{\delta
t}(\Cee \to \Cee')=1$.  Then, we can write (\ref{master1}) as
\begin{equation}
\label{master2}
P(\Cee, t+\delta t) - P(\Cee, t) = \sum_{\Cee' \ne \Cee} \left[
P(\Cee', t) M_{\delta t}(\Cee'\to\Cee) - P(\Cee, t) M_{\delta
t}(\Cee\to\Cee') \right]\;.
\end{equation}
The first term on the right-hand side of this equation gives the gain
in probability from transitions into the state $\Cee$, and the second
term the loss from transitions out of $\Cee$.  A steady state is
reached when $P(\Cee, t+\delta t) = P(\Cee, t) \equiv P^\ast(\Cee)$
and the \emph{total} loss and gain terms balance.  The very special
case where the losses and gains cancel term-by-term in (\ref{master2})
is called \emph{detailed balance} \cite{vanKampen92,Kelly79}, i.e.,
\begin{equation}
\label{db}
P^\ast(\Cee) M(\Cee\to\Cee') = P^\ast(\Cee^\prime) M(\Cee'\to\Cee) \;.
\end{equation}

This particular balancing of the loss and gain is required to satisfy
the time-reversal symmetry property (\ref{trsprop}).  This one sees
from the fact that (\ref{trsprop}) implies that in the steady state,
the probability of observing a particular trajectory $\Cee_0 \to
\Cee_1 \to \Cee_2 \cdots \to \Cee_N$ equals that of its time
reversal.  That is, we must have
\begin{eqnarray}
\label{trs}
P^{\ast}(\Cee_0) M_{\delta t}(\Cee_0 \to \Cee_1) M_{\delta t}(\Cee_1
\to \Cee_2) \cdots M_{\delta t}(\Cee_{N-1} \to \Cee_N) =\hspace{10em}
\nonumber\\ P^{\ast}(\Cee_N) M_{\delta t}(\Cee_N \to \Cee_{N-1})
M_{\delta t}(\Cee_{N-1} \to \Cee_{N-2}) \cdots M_{\delta t}(\Cee_{1}
\to \Cee_0)
\end{eqnarray}
for \emph{all} trajectories of any length $N$ \cite{Kelly79}.  The
special case $N=1$ reduces to the detailed balance condition
(\ref{db}). Note also that if one forms a loop in state space by
putting $\Cee_N=\Cee_0$ one obtains the \emph{Kolmogorov criterion}
\cite{Kelly79}
\begin{eqnarray}
\label{kc}
M_{\delta t}(\Cee_0 \to \Cee_1) M_{\delta t}(\Cee_1 \to \Cee_2) \cdots
M_{\delta t}(\Cee_{N-1} \to \Cee_N) =\hspace{10em} \nonumber\\ =
M_{\delta t}(\Cee_N \to \Cee_{N-1}) M_{\delta t}(\Cee_{N-1} \to
\Cee_{N-2}) \cdots M_{\delta t}(\Cee_{1} \to \Cee_0)
\end{eqnarray}
which is an equivalent statement of detailed balance.  This expression
is useful because often when formulating a stochastic dynamical model
(e.g., to describe nonequilibrium or complex systems as described in
the Introduction) one does not know the stationary distribution
$P^\ast(\Cee)$ in advance, only the transition probabilities
$M_{\delta t}(\Cee\to\Cee')$.  Since (\ref{kc}) contains only the
latter, one can establish whether or not the steady state is
time-reversal symmetric or not.  If it is, one can then find a
potential function $V(\Cee)$ that is uniquely defined in terms of the
differences
\begin{equation}
\label{pd}
V(\Cee') - V(\Cee) = \frac{1}{\beta} \ln \left[ \frac{M_{\delta
t}(\Cee'\to\Cee)}{M_{\delta t}(\Cee\to\Cee')} \right]
\end{equation}
for pairs of configurations between which transitions in a single
timestep occur.  In other words, summing these differences along any
path between two configurations always gives the same result, implying
that forces are conservative and that the stationary distribution is
Boltzmann: $P^\ast(\Cee) \propto \exp[-\beta V(\Cee)]$.

Conversely, in the absence of time reversal symmetry, a potential
function defined via (\ref{pd}) becomes multivalued: forces are then
nonconservative and the dynamics dissipate energy.  Furthermore,
(\ref{kc}) implies a circulation of probability in the steady state,
which are likely to be manifested as macroscopic currents.  These
decidedly nonequilibrium effects lead us to conclude that the defining
property of nonequilibrium dynamics is a lack time-reversal symmetry
in the steady state: i.e., dynamics for which none of the (equivalent)
conditions (\ref{db}), (\ref{trs}) and (\ref{kc}) hold.

\section{Phase transitions from a microscopic dynamical point of view}
\label{pts}

We now describe how a stochastic dynamics---assumed Markov,
time-translationally invariant and ergodic, but not necessarily
time-reversal symmetric---is drawn to particular regions of
configuration space.  Such considerations are paramount in Monte Carlo
simulation and so descriptions similar to the following can be found
in textbooks on the subject such as \cite{BH97}.

\begin{figure}
\begin{center}
\includegraphics[width=0.90\linewidth]{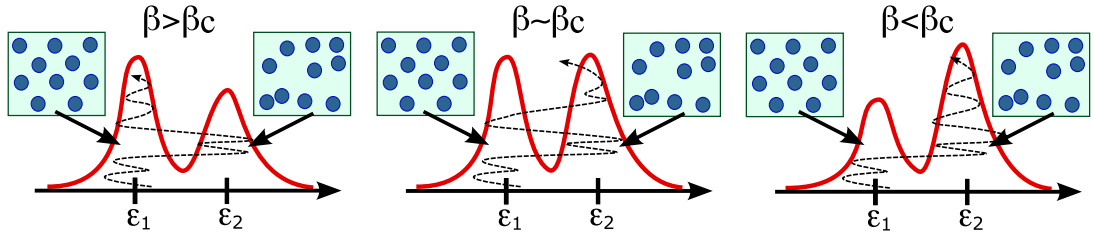}
\end{center}
\caption{Exploration of a doubly-peaked stationary distribution whose
  weight shifts between peaks as the inverse temperature $\beta$ is
  decreased.}
\label{AsymmetricPeaks}
\end{figure}

For concreteness, we base our discussion around a hypothetical,
doubly-peaked steady-state distribution $h(\epsilon)$ of a macroscopic
quantity $\epsilon$ which we notionally take to be an energy
density---see Fig.~\ref{AsymmetricPeaks}.  We shall associate the
lower-energy peak with an ordered (solid) phase, and the higher-energy
peak with a disordered (fluid) phase.  One can define the weight
(area) of the distribution under a peak as being proportional to the
time a dynamical trajectory---whether equilibrium or
nonequilibrium---spends under that peak, as the length of that
trajectory tends to infinity.  Equivalently (because the process is
ergodic \cite{Kelly79}), this distribution is given by the stationary
solution of the master equation (\ref{master1}).  Either way, a phase
transition occurs---roughly speaking---at a critical value of an
external parameter $\beta$ (notionally, inverse temperature) at which
peak containing the majority weight changes, as shown in
Fig.~\ref{AsymmetricPeaks}.

To be more precise, let us assume that for sufficiently large volumes
$V$, the statistical weight (\emph{unnormalised} probability)
$h(\epsilon)$ of the macrostate $\epsilon$ can be expressed in the
from
\begin{equation}
\label{height}
h(\epsilon) \approx \rme^{-\beta V g_{\beta}(\epsilon)}
\end{equation}
in which the function $g_{\beta}(\epsilon)$ is independent of $V$ but
changes with $\beta$.  We base this assumption on the fact that at
equilibrium, $g_{\beta}(\epsilon)$ is a free-energy density---there
is, however, no reason to assume that this statement is generally
true.  Nevertheless, we shall persevere with this assumption in order
to investigate the thermodynamic (infinite volume) limit.

Consider first the situation shown in Fig.~\ref{AsymmetricPeaks} where
the two phases corresponding to the peaks are not related by a
symmetry transformation (as is the case for a fluid and a solid).
Approximating each peak by a Gaussian, it is easy to show that the
ratio of weights under the peaks grows exponentially with volume as
$\exp(-\beta V[g_{\beta}(\epsilon_1) - g_{\beta}(\epsilon_2)])$.  Thus
the peak with the greater \emph{height}---or lowest free energy
$g(\epsilon)$---is overwhelmingly dominant in the thermodynamic limit.
Furthermore, this overwhelming dominance shifts \emph{suddenly} from
one region of state space to another at the temperature at which one
peak height overtakes the other.  It is this sudden change, described
mathematically by a sharp discontinuity in the limit of infinite
volume, that is the defining property of a phase transition.

At the transition point where both peaks have the same height, each
possesses a non-vanishing fraction of the total weight.  This
corresponds to the physical phenomenon of \emph{phase coexistence}
which, along with \emph{latent heat}, is a signature of an equilibrium
first-order transition.  Near the transition, where the difference in
peak heights is not too great, an initial condition under the smaller
peak can persist for long times indicating the presence of
\emph{metastability} which is also typically observed near a
first-order equilibrium phase transition.  Although the lifetime of
the metastable phase depends on the distance from the transition point
and the precise nature of the stochastic dynamical rules (whether
time-reversal symmetric or not), one expects by analogy with a Kramers
problem for the escape of a potential well through diffusion
\cite{Gardiner04,HTB90} that it will grow exponentially in the
system's volume.

\begin{figure}
\begin{center}
\includegraphics[width=0.9\linewidth]{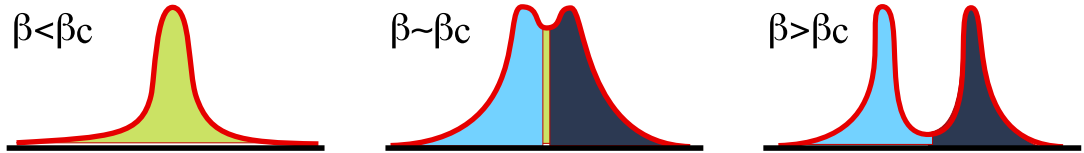}
\end{center}
\caption{\label{SymmetricPeaks} Splitting of a peak in the stationary
    distribution at a symmetry-breaking transition.}
\end{figure}

A different situation occurs at a transition at which a single peak
splits into two (or more) \emph{symmetric} peaks as shown in
Fig.~\ref{SymmetricPeaks}. This occurs, for example, if a paramagnet
spontaneously orders as the temperature is lowered.  The symmetry
implies that for a whole range of temperatures (not just at a single
point), each peak has precisely the same area.  Although, given long
enough, an ergodic dynamics would then spend the same amount of time
in each phase, it turns out again that the time taken to surmount the
barrier between the phases grows exponentially with the volume and so
in the thermodynamic limit, the dynamics only get to explore the peak
under which they start.  This describes a symmetry-breaking
transition.  At equilibrium such \emph{continuous} transitions are
typically accompanied by diverging correlation lengths and times.

In this description of phase transitions we have assumed only a
general property of peaks in the stationary distribution and
ergodic---but not necessarily equilibrium---dynamics.  However, it is
worth for a moment specialising to the equilibrium case where much
more is known.  As noted above, the function $g(\epsilon)$ is then a
free energy that the dynamics tend to minimise.  Macroscopically, one
can often ascribe a phase transition to a competition between
minimising the energy and maximising the entropy, the relative balance
of the two controlled by the temperature.  It is interesting to
observe that this competition is also a feature of a stochastic
equilibrium dynamics.

Let us consider as a concrete example the \emph{Metropolis} algorithm
which is popular in Monte Carlo simulation as a method to generate a
sequence of configurations that samples an equilibrium distribution
\cite{NB99,Janke}.  It is implemented by repeatedly iterating two
steps:
\begin{enumerate}
\item\label{m1} Given a configuration $\Cee$, one generates a trial
  subsequent configuration $\Cee'$ using a stochastic rule that, on
  its own, is time-reversal symmetric with respect to a uniform
  distribution over state space.  For example, in an Ising model, one
  could flip a randomly-chosen spin.
\item\label{m2} If $\Cee'$ has a lower energy than $\Cee$, it is
  always used as the next configuration in the sequence.  Otherwise,
  it is retained only with a probability $\rme^{-\beta
  [E(\Cee')-E(\Cee)]}$.
\end{enumerate}
From the discussion of the previous section, it is straightforward to
show that these dynamics generate time-reversal symmetric trajectories
sampling the Boltzmann distribution $P^{\ast}(\Cee) \propto
\rme^{-\beta E(\Cee)}$.  Notice particularly how the energy-entropy
competition arises: step (\ref{m1}) generates an attraction towards
high-entropy macrostates (i.e., those that are realised by a large
number of microstates) whilst step (\ref{m2}) repels those
high-entropy states that involve too great an increase in energy.

\section{Mathematical characterisation of discontinuities at a phase
  transition}
\label{math}

In the previous section we found that under the assumption of the
statistical weight of macrostates varying with system volume $V$
according to (\ref{height}), an overwhelmingly large proportion of the
weight resides under the peak with the greater height in the limit of
infinite $V$.  Therefore, the normalisation of this weight function
\begin{equation}
Z(\beta) = \int \rmd\epsilon h(\epsilon) \approx \int \rmd\epsilon
\rme^{-\beta V g_{\beta}(\epsilon)}
\end{equation}
closely approximates the area under the highest peak (or peaks when
more than one have the same height).  Approximating once again the
peaks as Gaussians one finds that
\begin{equation}
\label{feq}
f(\beta) = -\lim_{V\to\infty} \frac{\ln Z(\beta)}{\beta V} =
\min_{\epsilon} \{ g_{\beta}(\epsilon) \}
\end{equation}
gives the free-energy-like quantity corresponding to the highest
peak(s).  At equilibrium, $f(\beta)$ is the equilibrium (Helmholtz)
free energy density and, as is well understood (see e.g.,
\cite{Stanley87}) signals a phase transition through a nonanalyticity
in $f(\beta)$.  In particular, a discontinuous first derivative
implies a jump in a quantity like the energy or density of the system,
i.e., a first-order transition, whilst a discontinuous higher
derivative corresponds to a continuous transition.

Under the assumption that the peaks of a stationary distribution
arising from nonequilibrium dynamics behave according to
(\ref{height}), we can always characterise the height of the largest
peak using the \emph{normalisation} of the distribution $Z(\beta)$ in
exactly the same way as we do for an equilibrium distribution to find
the equilibrium free energy.  What is missing is a general expression
for the normalisation and any association between particular
nonanalyticities in the resulting free energy (\ref{feq}) and
macroscopically observable behaviour.  In the remainder of this
article we address these points.

First, we show that it is possible to write down a general expression
for the normalisation $Z(\beta)$ (see also \cite{EB02}).  This is
achieved by casting the master equation (\ref{master2}) as a matrix
equation
\begin{equation}
\vec{P}(t+\delta t) - \vec{P}(t) = \mat{M} \vec{P}(t)
\end{equation}
in which the $i^{\rm th}$ element of $\vec{P}(t)$ gives the
probability for the system to be in configuration $\Cee_i$ at time
$t$, the off-diagonal matrix elements $[\mat{M}]_{ij} = M_{\delta
t}(\Cee_j \to \Cee_i)$ encode the gain terms and the diagonal matrix
elements the loss terms $[\mat{M}]_{ii} = - \sum_j M_{\delta t}(\Cee_i
\to \Cee_j)$.  After imposing a normalisation constraint, Cramer's
rule \cite{LT85} can be employed to solve the matrix equation $\mat{M}
\vec{P}^\ast = 0$ for the stationary distribution $\vec{P}^\ast$.
This procedure gives
\begin{equation}
\label{cramer}
P^\ast(\Cee_i) = \frac{ \det \widetilde{\mat{M}}^{(i)} }{Z}
\quad\mbox{with}\quad Z = \sum_i \det \widetilde{\mat{M}}^{(i)} .
\end{equation}
where $\widetilde{\mat{M}}^{(i)}$ is the matrix obtained by striking
out both row and column $i$ of $\mat{M}$.  We shall use the
normalisation $Z$ that arises through this procedure to define a
generalised partition function for any system with a unique steady
state, whether equilibrium or nonequilibrium.

This form of $Z$ implies, via the matrix tree theorem
\cite{Harary69,BlythePhD,BdGR04}, that it is the partition function
for an ensemble of spanning in-trees on the graph of transition
probabilities with a particular edge weighted by its corresponding
transition probability $M(\Cee\to\Cee')$.  This suggests an
interpretation (albeit abstract) of the transition probabilities as
equilibrium fugacities: we shall see this more concretely later.  In
the meantime, we note that a drawback of (\ref{cramer}) is that
calculation of the determinants is intractable for a model with even a
modest number of microstates, let alone in the thermodynamic limit of
infinite volume which is required for a free energy to develop a
nonanalyticity, and thereby predict a phase transition.

One way to gauge the possibility of a phase transition in an
infinite system when one only has finite-size results to hand is to
look at the \emph{zeros} of the partition function in the complex
plane of external parameters (in this context, the transition rates),
an approach that was introduced by Lee and Yang in the 1950s
\cite{YL52,LY52}. The historical development of these ideas, and their
application to both equilibrium and nonequilibrium steady states, has
recently been very thoroughly reviewed in \cite{BDL05}, an effort that
extends our own overview of the nonequilibrium cases in \cite{BE03}.
Therefore, we cover only the most salient points here.

For simplicity, let us assume that every nonzero transition
probability is some integer power of an external parameter $z$ (e.g.,
in an Ising model one might have $z=\rme^{-\beta}$).
Eq.~(\ref{cramer}) then implies that the partition function $Z$ is a
polynomial in $z$, its zeros lying off the positive real axis in the
complex-$z$ plane because $Z$ is a sum of positive statistical
weights.  This means that the free energy, obtained from the partition
function using (\ref{feq}), has only \emph{isolated} logarithmic
singularities in the complex plane, and hence is analytic for the
entire range of $z$ that can be accessed physically (i.e., the
positive real $z$ axis).

\begin{figure}
\begin{center}
\includegraphics[scale=0.5]{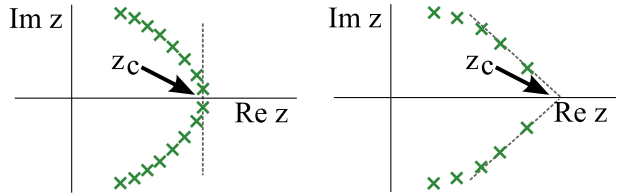}
\end{center}
\caption{\label{ZerosLoci} Loci of partition function zeros
characteristic of a (left) first-order and (right) continuous phase
transition at $z_c$ as described in the text.}
\end{figure}

The only exception to this is if zeros approach a point $z_c$ on the
positive real axis arbitrarily closely in the thermodynamic limit.  It
turns out that typically the zeros approach $z_c$ along a line $\Re
f_1(z) = \Re f_2(z)$ separating two different analytic free energies
$f_1(z)$ and $f_2(z)$.  By expanding the free energy difference
between the two phases near the transition point $z_c$, one finds (see
\cite{BDL05,BE03} for details and original references) that if the
free energy has a discontinuity in its first derivative, as it would
at an equilibrium first-order transition, the line of zeros passes
smoothly along a parabola through $z_c$ and with a non-vanishing
density of zeros at that point.  At a point where a higher derivative
is discontinuous (as occurs at an equilibrium continuous transition)
the density of zeros vanishes as the critical point is approached, the
path forming a cusp.  These characteristics are indicated
schematically in Fig.~\ref{ZerosLoci}.  Thus we see that the zeros of
the partition function give a nice way to represent nonanalyticities
in the associated free energy which, in turn, relate to structural
changes in peaks of the stationary distribution as described in
Section~\ref{pts}.  Recall, however, that we have no general theory to
suggest that, for example, a discontinuity in the first derivative of
the free energy necessarily implies that the physical phenomena
associated with a first-order transition will be observed in a
nonequilibrium steady state.  This possibility we now explore with
reference to an exactly solved model.

\section{Case study: The asymmetric exclusion process with open boundaries}

\begin{figure}
\begin{center}
\includegraphics[width=0.6\linewidth]{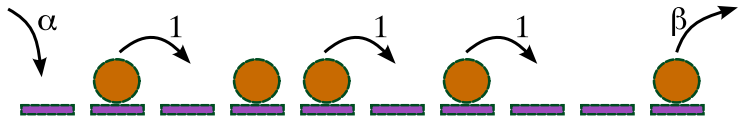}
\end{center}
\caption{Allowed microscopic transitions in the ASEP and
  the associated transition rates.}
\label{asep} 
\end{figure}

The asymmetric exclusion process (ASEP) with open boundaries is an
exemplar of an interacting particle system that is driven by its
environment.  Introduced as a model of biopolymerisation \cite{MGP68},
and relevant also to the study of traffic flow \cite{Schadschneider01}
and biological transport \cite{CSN05}, it comprises a one-dimensional
lattice of $L$ sites, each of which may be occupied by at most one
particle by virtue of a hard-core exclusion constraint.  In a timestep
of duration $\delta t$, a particle in the bulk hops one site to the
right with probability $\delta t$ as long as the hard-core exclusion
constraint is not violated as a result.  Additionally, the left-most
site can---if empty---be populated with a particle with probability
$\alpha \delta t$ and a particle at the rightmost site removed with
probability $\beta \delta t$.  These transitions are illustrated in
Fig.~\ref{asep}.  We consider the continuous-time limit $\delta t\to0$
so that at most one event occurs in a timestep: this defines a
\emph{random-sequential} updating scheme in which $\alpha$ and $\beta$
are transition rates.  The fact that no elementary particle move can
immediately be reversed implies a lack of time-reversal symmetry in
the dynamics.  Particularly, there is a nonzero particle current in
its nonequilibrium steady state.

\begin{figure}
\begin{center}
\includegraphics[width=0.9\linewidth]{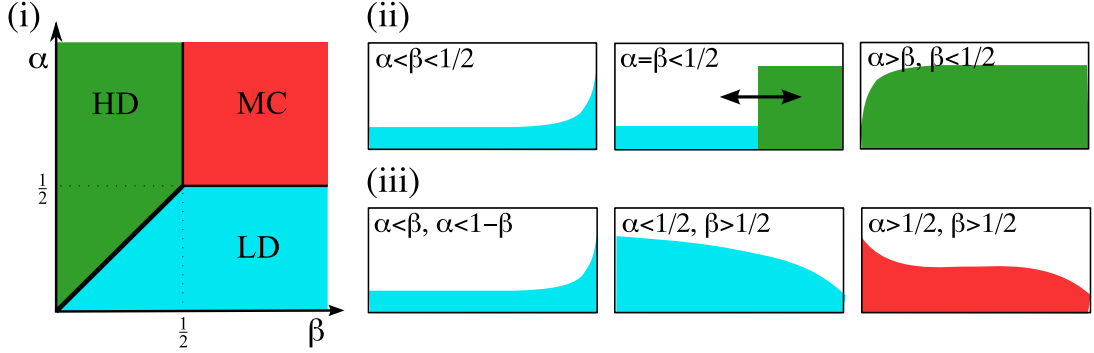}
\end{center}
\caption{\label{AsepPhaseBehaviour} Phase behaviour of the ASEP: (i)
  the phase diagram; (ii) density profiles across the transition
  between the low- and high-density phases; (iii) density profiles
  across the transition between the low-density and maximal current
  phases.}
\end{figure}

The main reason for discussing the model in the present work is that
its steady-state distribution has been calculated exactly through two
different (but equivalent) approaches \cite{SD93,DEHP93}.  In
particular this means such properties as the partition function, free
energy, peaks in the distribution and so on can be extracted and
compared with the macroscopic phase behaviour present in the model.
There are, in fact, three distinct phases in the steady state:
\begin{enumerate}
\item \textit{High-density (HD) phase} --- When the exit rate $\beta$
  is lower than the entry rate $\alpha$ and further smaller than a
  critical value of $\frac{1}{2}$, particles accumulate at the right
  boundary creating a region of high density $\rho=1-\beta$ that
  extends into the bulk, before decaying exponentially towards the
  left boundary.  The slow exit rate further limits the current to
  $J=\beta(1-\beta)$.
\item \textit{Low-density (LD) phase} --- Since the system is
  invariant under the combined transformation of a particle-hole
  exchange and left-right reflection of the lattice, it follows that
  for $\alpha<\beta, \alpha<\frac{1}{2}$ one has a transformed version
  of the HD phase.  This has a low density $\rho=\alpha$ extending
  from the left boundary into the bulk along with an exponential decay
  from the right boundary and a current $J=\alpha(1-\alpha)$.
\item \textit{Maximal-current (MC) phase} --- When both entry and exit
  rates $\alpha$ and $\beta$ are greater than $\frac{1}{2}$, the
  boundaries no longer limit the current and instead a particle's
  progress is impeded by the hard-core exclusion.  In this phase the
  current assumes a constant value $J=\frac{1}{4}$, the maximum for
  any combination of $\alpha$ and $\beta$.  The density is
  $\rho=\frac{1}{2}$ in the bulk and has a power-law decay with
  exponent $\frac{1}{2}$ both from the left and towards the right
  boundary.
\end{enumerate}
The phase diagram is shown in Fig.~\ref{AsepPhaseBehaviour}~(i).

We are particularly interested in the physics near the transition
points.  Along the boundary between the HD and LD phases,
$\alpha=\beta<\frac{1}{2}$, there is phase coexistence with low- and
high-density regions separated by a shock front that performs a random
walk along the lattice \cite{ABL88}.  This phenomenology, shown in
Fig.~\ref{AsepPhaseBehaviour}~(ii), is reminiscent of an equilibrium
first-order transition.  Meanwhile as the MC phase is approached from
either the LD or HD phase, the lengthscale characterising the density
decay diverges, as often occurs at a continuous equilibrium phase
transition but with power-law correlations exhibited generically in
the MC phase.  See Fig.~\ref{AsepPhaseBehaviour}~(iii).

\begin{figure}
\begin{center}
\includegraphics[scale=0.33]{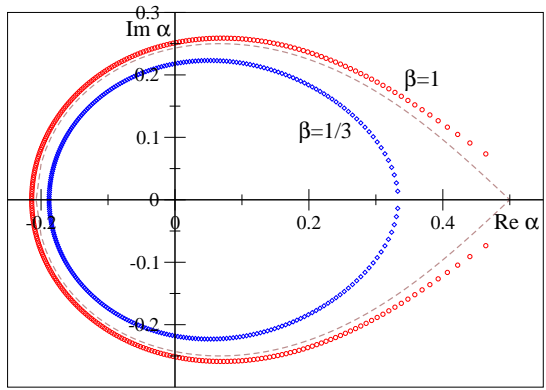}
\end{center}
\caption{\label{AsepZeros} Zeros of the ASEP's partition function
(\ref{Zasep}) for a lattice of $L=300$ sites in the complex-$\alpha$
plane.  For $\beta=\frac{1}{3}$, a first-order phase transition occurs
at $\alpha=\frac{1}{3}$. At $\beta=1$ a continuous transition at
$\alpha=\frac{1}{2}$ occurs.  The dashed line shows the phase-boundary
determined analytically \cite{BE02} for the latter case.}
\end{figure}

The partition function (\ref{cramer}) for an $L$-site lattice is known
from the exact solutions \cite{SD93,DEHP93} to be \footnote{Note that
the expression in \cite{DEHP93} omits an unimportant factor of
$(\alpha\beta)^L$ which is present if one follows through the solution
employing matrix determinants (\ref{cramer}).}
\begin{equation}
\label{Zasep}
Z_L = \sum_{n=0}^{L} \frac{(L-n)(L+n-1)!}{L!n!} (\alpha\beta)^{n}
\sum_{r=0}^{L-n} \alpha^r \beta^{L-n-r} \;.
\end{equation}
We explore the nonanalyticities in the free energy (\ref{feq}) by
examining the zeros of this partition function for a finite system, as
described in the previous section.  Here, we go into the
complex-$\alpha$ plane at fixed $\beta$ (without loss of generality,
because of the symmetry under the exchange $\alpha \leftrightarrow
\beta$ and of particles and holes).  In Fig.~\ref{AsepZeros},
numerical solutions for the zeros with $\beta=\frac{1}{3}$ and
$\beta=1$, both with $L=300$ are shown.  In the former case the path
of zeros passes smoothly through a point on the real axis with
$\alpha\approx\frac{1}{3}$.  Meanwhile, in the latter case we have a
path of zeros meeting at an angle of $\pi/2$.  As previously
described, these patterns are characteristic of equilibrium
first-order and continuous phase transitions and the physical
phenomena seen at the transitions (see above) correspond with these
classifications.  This result is quite intriguing given that the
logarithm of (\ref{Zasep}) is not an equilibrium free energy, nor are
$\alpha$ and $\beta$ equilibrium fugacities.  One can attempt to
explain this result by noting that the free energy and current are
simply related \cite{DEHP93} via $f(\alpha,\beta) = \ln
J(\alpha,\beta) - \ln \alpha\beta$, a fact that allows the locus of
partition function zeros to be determined analytically \cite{BE02}.
However, this explanation amounts to a statement
that the correct order parameters to characterise the phase
transitions are the derivatives of $J$ with respect to $\alpha$ and
$\beta$---a statement which I do not find obvious.

To turn full circle, I conclude by examining peaks in the ASEP's
stationary distribution.  It is possible, though technically
difficult, to determine the distribution over the time-integrated
current \cite{BD05}, instantaneous current-density \cite{DS04} and
even spatially-dependent density profile \cite{DLS03} microstates.
Here, we take a short-cut which brings additional insight in terms of
an energy-entropy competition at the expense of losing sight of the
dynamics.

\begin{figure}
\begin{center}
\includegraphics[width=0.6\linewidth]{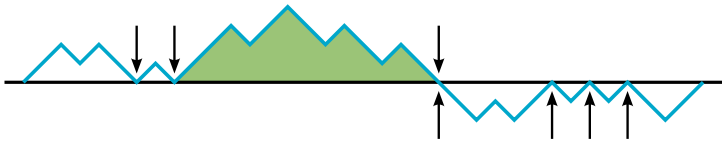}
\end{center}
\caption{\label{Surface} An equilibrium surface whose partition
  function coincides with that for the ASEP.  The arrows indicate
  contacts with the origin which lower the surface's energy whilst
  entropy is gained from excursions away from the origin, such as that
  shown shaded.}
\end{figure}

The idea is to construct an \emph{equilibrium} statistical mechanical
model whose partition function coincides with (\ref{Zasep}).  One
prescription \cite{BdGR04,BJJK04a} is a surface on a one-dimensional
surface whose left portion is constrained to lie above an origin and
its right portion below---see Fig.~\ref{Surface}.  Contacts with the
origin from above (respectively, below) contribute an energy $\ln
\alpha$ ($\ln\beta$).  Note that $\alpha$ and $\beta$ are equilibrium
fugacities in this interpretation, and when sufficiently small allow
the lowering of the energy by contact with the origin to beat the
entropy gained from large excursions of the surface from the origin.
This competition gives rise to three equilibrium phases: one in which
the surface is adsorbed from above, one from below and a third in
which it is completely desorbed.  These correspond to the ASEP's LD,
HD and MC phases respectively, and the phase diagram is exactly the
same as in Fig.~\ref{AsepPhaseBehaviour}~(i) (as it must be).
Furthermore, the preceding classification of the transitions is
uncontroversial in this picture since derivatives of the free energy
with respect to $\alpha$ and $\beta$ give the appropriate order
parameters for the model, viz, the density $\rho_a$ of contacts with
the origin from above and $\rho_b$ from below.

\begin{figure}
\begin{center}
\includegraphics[width=0.3\linewidth]{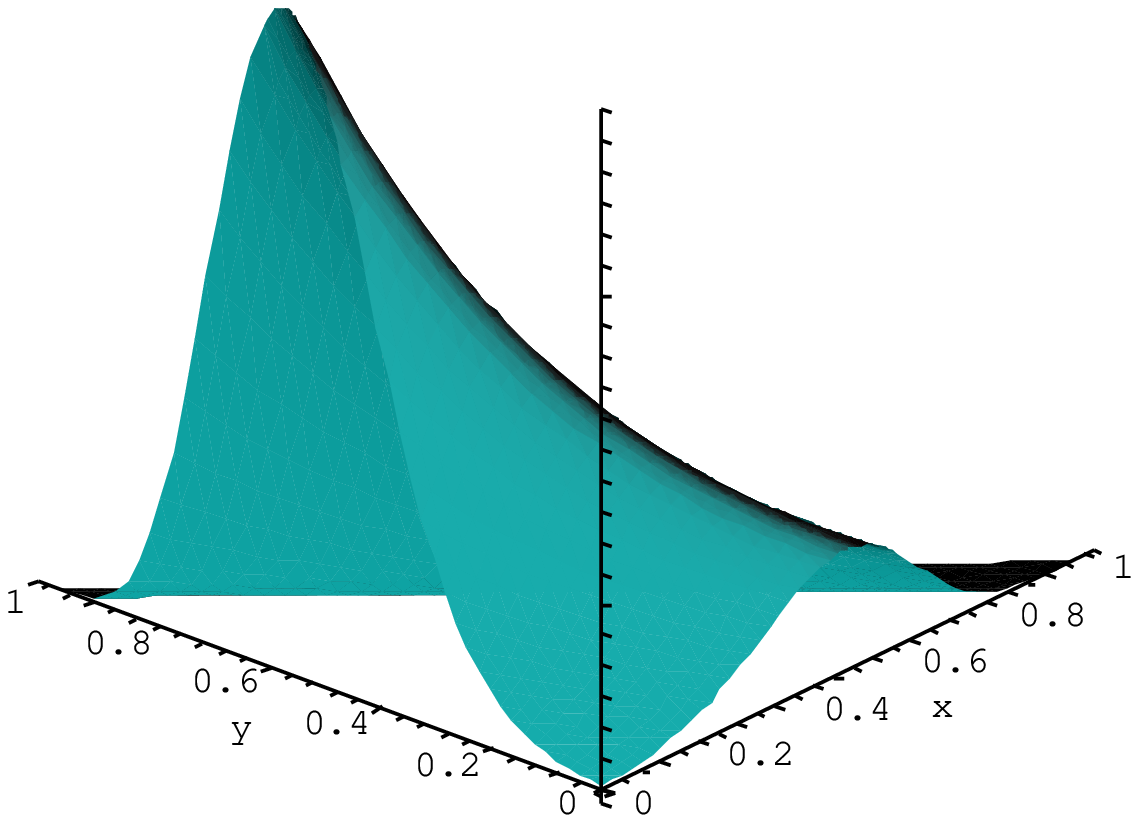}
\includegraphics[width=0.3\linewidth]{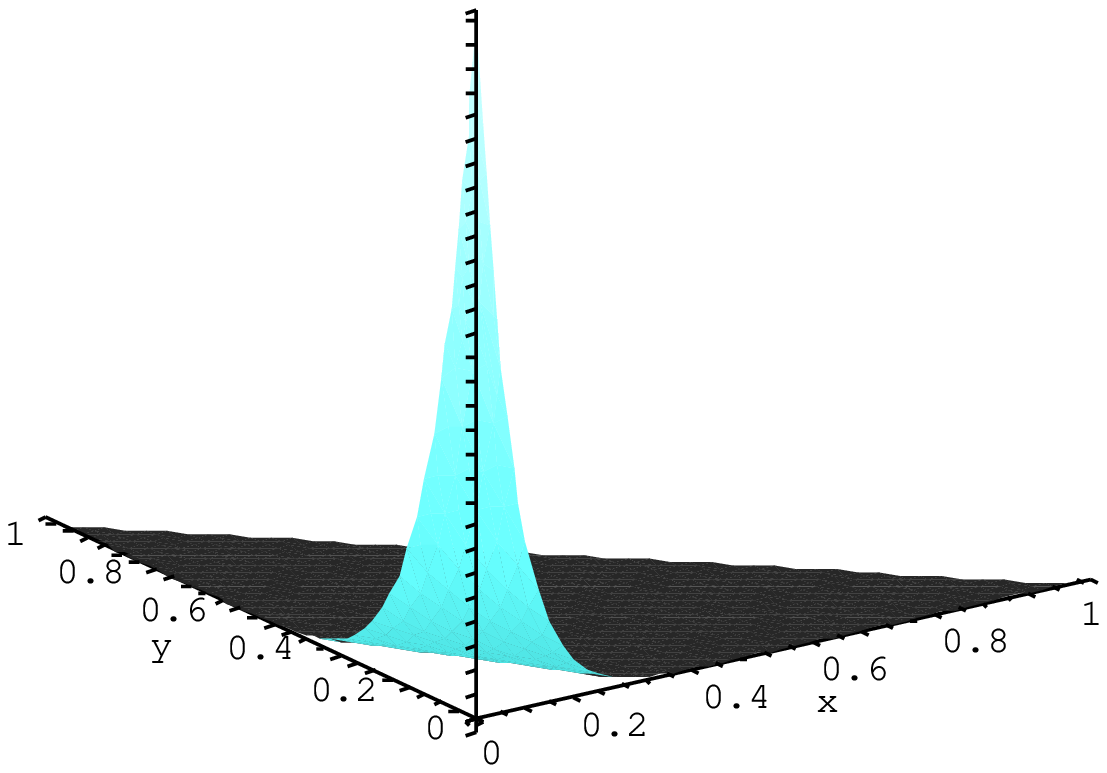}
\includegraphics[width=0.3\linewidth]{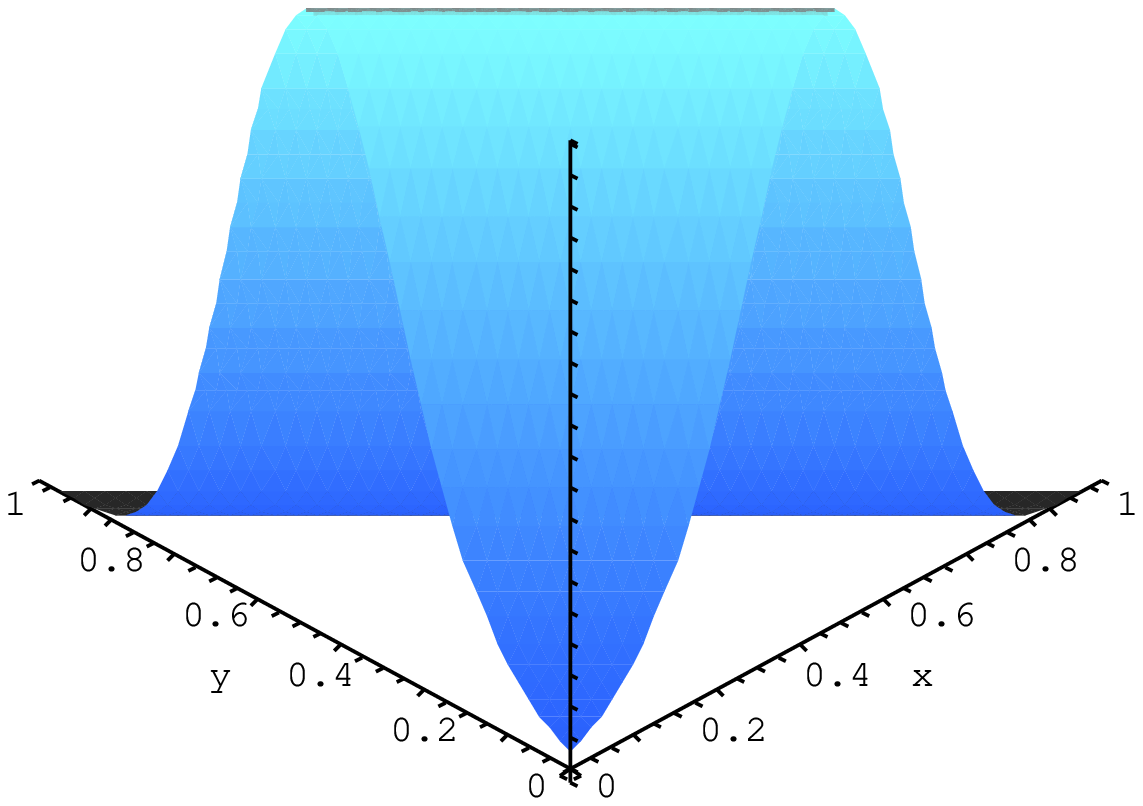}
\end{center}
\caption{\label{ASEPpeaks} Peaks in the $\rho_a$-$\rho_b$ plane
(labelled $x$ and $y$ respectively in the figures) for a large, but
finite, surface length $L$.  From left to right: $\alpha=\frac{1}{3},
\beta=\frac{3}{10}$; $\alpha=\frac{7}{8}, \beta=\frac{2}{3}$; and
$\alpha=\beta=\frac{1}{3}$.}
\end{figure}

In the space spanned by these order parameters, the stationary
distribution is \cite{BdGR04,BJJK04b}
\begin{equation}
h(\rho_a, \rho_b) \sim \left[ \frac{
(2-\rho_a-\rho_b)^{2-\rho_a-\rho_b} }{ \alpha^{\rho_a} \beta^{\rho_b}
(1-\rho_a-\rho_b)^{1-\rho_a-\rho_b} } \right]^L
\end{equation}
which has the exponential growth with system size $L$ prescribed in
Section~\ref{pts}.  As can be seen from Fig.~\ref{ASEPpeaks}, the
three equilibrium phases correspond to peaks on the boundary of the
physical region $0 \le \rho_a+\rho_b \le 1$.  In the HD phase
(adsorption from above), the dominant peak is at a point with
$\rho_a=0, \rho_b>0$, whilst the peak corresponding to the LD phase
(adsorption from below) on the line $\rho_b=0$ in exponentially
suppressed.  In the entropy-dominated desorbed phase, these two peaks
merge into one at $\rho_a=\rho_b=0$.  This is reminiscent of the
discussion of the continuous phase transition in Section~\ref{pts},
although in the present case no symmetry breaking is involved.  At the
first-order transition, the situation is a little different to that
described in Section~\ref{pts}.  Rather than two isolated peaks of
equal height, we have a continuous ridge linking the LD and HD phase
peaks.  That is, any mixture of these two phases is represented with
equal weight in this distribution, a fact that is reflected in the
diffusive wandering of the shock front separating them in the ASEP
(discussed above): indeed, one also finds this feature in the
distribution in the density-current plane \cite{DS04}.  Thus we see
that, broadly speaking at least, the general features outlined in
Section~\ref{pts} are present in this particular nonequilibrium
distribution.

\section{Conclusion}

In this article, I have tried to identify physical and
mathematical properties that might be shared by both equilibrium and
nonequilibrium phase transitions, the latter being defined with
reference to stationary states generated by dynamics that are not
time-reversal symmetric.  In equilibrium systems, we understand the
importance of a competition between energy and entropy in shaping the
phase behaviour; away from equilibrium, the proper macroscopic
characterisation is unclear.  However, as I have shown, when the
equilibrium theory of phase transitions is presented from the
perspective of microscopic stochastic dynamics, one has a picture that
applies equally well to nonequilibrium dynamical systems as long as
two criteria are satisfied.  First, the dynamics must be ergodic---a
property held by many nonequilibrium dynamical models (but see below).
We also require a macroscopic phase to be associated with a peak in
the stationary distribution that grows in a prescribed way with the
system's volume.  Some additional precision in the definition of this
association and clarity regarding its necessity is required.

Nevertheless, quantitative results relating to the well-studied
asymmetric exclusion process (ASEP) are encouraging.  As shown here,
the stationary distribution in the space of macroscopic order
parameters of a related equilibrium model behaves roughly as described
in Section~\ref{pts}.  A study elsewhere \cite{DS04} shows similar
structures in the space of nonequilibrium current and density
macrostates, indicating a robustness that is not immediately obvious.
Another curiosity is that nonanalyticities in a rather
abstractly-defined free energy reflect the physical nature of phase
transitions, as evidenced by patterns of partition function zeros.
This is not an isolated case, similar results having been established
for a number of related models \cite{BE02,BDL05}.  Although this
suggests an intimate relationship between equilibrium and
nonequilibrium phase transitions, one should not rule out the
possibility that these models have been amenable to such detailed
study precisely \emph{because} they possess this special feature.

Finally, we should outline some criticisms of the theory outlined in
Section~\ref{math}. Most obviously, the expression (\ref{cramer})
seems overprescribed: if one has an equilibrium dynamics, evaluation
of the determinant in (\ref{cramer}) is likely to give a much more
complicated expression than that obtained by summation of the
Boltzmann factors.  Although one does not expect such additional
factors to contribute any nonanalyticities to the equilibrium free
energy, there is a problem demonstrated by certain spin glass models
\cite{CC05}.  Specifically, at a dynamic transition temperature, the lifetime of a metastable state diverges, precluding relaxation
to the paramagnetic state that has a lower free energy.  Since $Z$ as given here in Eq.~(\ref{cramer})
can be shown \cite{EB02,BlythePhD} to be inverse to the product of
relaxation times, one might expect nonanalyticities in $Z$ unrelated
to static phase transitions to arise after all. More widely, the
notion of a partition function in the presence of such ergodicity
breaking is at best ill-defined.

To conclude with a more general remark, it is worth noting that in our
effort to understand nonequilibrium dynamics, it has become
traditional to specify \emph{ad hoc} stochastic rules with little
attention paid to such thermodynamical considerations as heat and
work.  A bridge between the two is provided by fluctuation theorems
\cite{Kurchan,ES02} such as Crooks' equality \cite{Crooks00} that
relates products of transition probabilities to dissipated work.  This
suggests that it might be possible to separate the potential
differences (\ref{pd}) for an arbitrary process into conservative and
nonconservative parts.  However, since the distinction is to some
extent arbitrary \cite{LS99}, it is unclear how physically meaningful
such a procedure would be.

\ack

I thank Martin Greenall, Andy Jackson and Jamie Wood for comments on
the manuscript, and the Royal Society of Edinburgh for the support of
a Personal Research Fellowship.


\vspace{2ex}

\begin{thebibliography}{10}

\bibitem{AGT05}
M~Henkel, M~Pleimling, and R~Sanctuary, editors.
\newblock {\em Ageing and the Glass Transition}.
\newblock Springer Verlag, 2005.
\newblock In press.

\bibitem{Kob}
W~Kob.
\newblock Introduction to the physics of glasses: From experiments to computer
  simulation.
\newblock In \cite{AGT05}.

\bibitem{Kruger}
J~Kr\"uger.
\newblock Experiments on the glass transition.
\newblock In \cite{AGT05}.

\bibitem{DS01}
P~G Debenedetti and F~H Stillinger.
\newblock Supercooled liquids and the glass transition.
\newblock {\em Nature}, 410:259, 2001.

\bibitem{Angell95}
C~A Angell.
\newblock Formation of glasses from liquids and biopolymers.
\newblock {\em Science}, 267:1924, 1995.

\bibitem{BCKM97}
J-P Bouchaud, L~F Cugliandolo, J~Kurchan, and M~M\'ezard.
\newblock Out of equilibrium dynamics in spin-glasses and other glassy systems.
\newblock In \cite{Young97}.

\bibitem{Young97}
A~P Young, editor.
\newblock {\em Spin glasses and random fields}.
\newblock World Scientific, Singapore, 1997.

\bibitem{Vincent}
E~Vincent.
\newblock Aging, rejuvenation and memory: the example of spin glasses.
\newblock In \cite{AGT05}.

\bibitem{BarYam97}
Y~Bar-Yam.
\newblock {\em Dynamics of complex systems}.
\newblock Addison-Wesley, Reading, Mass., 1997.

\bibitem{vanKampen92}
N~G van Kampen.
\newblock {\em Stochastic Processes in Physics and Chemistry}.
\newblock Elsevier, Amsterdam, 1992.

\bibitem{Zaslavsky05}
G~M Zaslavsky.
\newblock {\em Hamiltonian Chaos and Fractional Dynamics}.
\newblock OUP, Oxford, 2005.

\bibitem{Kelly79}
F~P Kelly.
\newblock {\em Reversibility and stochastic networks}.
\newblock Wiley, New York, 1979.

\bibitem{BH97}
K~Binder and D~W Heermann.
\newblock {\em Monte Carlo Simulation in Statistical Physics}.
\newblock Solid-State Sciences. Springer Verlag, Berlin, 1997.

\bibitem{Gardiner04}
G~W Gardiner.
\newblock {\em Handbook of stochastic methods}.
\newblock Synergetics. Springer Verlag, Berlin, 3rd edition, 2004.

\bibitem{HTB90}
P~H\"anngi, P~Talkner, and M~Borkovec.
\newblock Reaction-rate theory: fifty years after {Kramer}.
\newblock {\em Rev. Mod. Phys.}, 62:251, 1990.

\bibitem{NB99}
M~E~J Newman and G~T Barkema.
\newblock {\em Monte Carlo methods in statistical physics}.
\newblock Clarendon Press, Oxford, 1999.

\bibitem{Janke}
W~Janke.
\newblock Introduction to simulation techniques.
\newblock In \cite{AGT05}.

\bibitem{Stanley87}
H~E Stanley.
\newblock {\em Introduction to phase transitions and critical phenomena}.
\newblock OUP, Oxford, 1987.

\bibitem{EB02}
M~R Evans and R~A Blythe.
\newblock Nonequilibrium dynamics in low-dimensional systems.
\newblock {\em Physica A}, 313:110, 2002.

\bibitem{LT85}
P~Lancaster and M~Tismenetsky.
\newblock {\em The theory of matrices}.
\newblock Academic Press, San Diego, 2nd edition, 1985.

\bibitem{Harary69}
F~Harary.
\newblock {\em Graph theory}.
\newblock Addison-Wesley, Reading, Mass, 1969.

\bibitem{BlythePhD}
R~A Blythe.
\newblock {\em Nonequilibrium steady states and dynamical scaling regimes}.
\newblock PhD thesis, University of Edinburgh, 2001.
\newblock Available from {http://www.ph.ed.ac.uk/cmatter.links/rab-thesis}.

\bibitem{BdGR04}
R~Brak, J~de~Gier, and V~Rittenberg.
\newblock Nonequilibrium stationary states and equilibrium models with long
  range interactions.
\newblock {\em J.\ Phys.\ A: Math. Gen.}, 37:4303, 2004.

\bibitem{YL52}
C~N Yang and T~D Lee.
\newblock Statistical theory of equations of state and phase transitions i:
  Theory of condensation.
\newblock {\em Phys. Rev.}, 87:404, 1952.

\bibitem{LY52}
T~D Lee and C~N Yang.
\newblock Statistical theory of equations of state and phase transitions ii:
  Lattice gas and ising model.
\newblock {\em Phys. Rev.}, 87:410, 1952.

\bibitem{BDL05}
I~Bena, M~Droz, and A~Lipowski.
\newblock Statistical mechanics of equilibrium and nonequilibrium phase
  transitions: The {Yang-Lee} formalism.
\newblock {\em Int.\ J.\ Mod.\ Phys.\ B}, 2005.
\newblock To appear; see also cond-mat/0510278.

\bibitem{BE03}
R~A Blythe and M~R Evans.
\newblock The {Lee-Yang} theory of equilibrium and nonequilibrium phase
  transitions.
\newblock {\em Braz. J. Phys.}, 33:464, 2003.

\bibitem{MGP68}
C~T MacDonald, J~H Gibbs, and A~C Pipkin.
\newblock Kinetics of biopolymerization on nucleic acid templates.
\newblock {\em Biopolymers}, 6:1, 1968.

\bibitem{Schadschneider01}
A~Schadschneider.
\newblock Statistical mechanics of traffic flow.
\newblock {\em Physica A}, 285:101, 2001.

\bibitem{CSN05}
D~Chowdhury, A~Schadschneider, and K~Nishinari.
\newblock Physics of transport and traffic phenomena in biology: from molecular
  motors and cells to organisms.
\newblock {\em Preprint}, 2005.
\newblock {physics/0509025}.

\bibitem{SD93}
G~Sch\"utz and E~Domany.
\newblock Phase transitions in an exactly soluble one-dimensional exclusion
  process.
\newblock {\em J. Stat. Phys.}, 72:277, 1993.

\bibitem{DEHP93}
B~Derrida, M~R Evans, V~Hakim, and V~Pasquier.
\newblock Exact solution of a 1d asymmetric exclusion model using a matrix
  formulation.
\newblock {\em J. Phys. A: Math. Gen.}, 26:1493, 1993.

\bibitem{ABL88}
E~D Andjel, M~Bramson, and T~M Liggett.
\newblock Shocks in the asymmetric simple exclusion process.
\newblock {\em Prob. Theory Rel. Fields}, 78:231, 1988.

\bibitem{BE02}
R~A Blythe and M~R Evans.
\newblock {Lee-Yang} zeros and phase transitions in nonequilibrium steady
  states.
\newblock {\em Phys. Rev. Lett.}, 89:080601, 2002.

\bibitem{BD05}
T~Bodineau and B~Derrida.
\newblock Current large deviations for asymmetric exclusion processes with open
  boundaries.
\newblock {\em Preprint}, 2005.
\newblock {cond-mat/0509179}.

\bibitem{DS04}
M~Depken and R~Stinchcombe.
\newblock Exact joint density-current probability function for the asymmetric
  exclusion process.
\newblock {\em Phys. Rev. Lett.}, 93:040602, 2004.

\bibitem{DLS03}
B~Derrida, J~L Lebowitz, and E~R Speer.
\newblock Exact large deviation functional of a stationary open driven
  diffusive system: the asymmetric exclusion process.
\newblock {\em J. Stat. Phys.}, 110:775, 2003.

\bibitem{BJJK04a}
R~A Blythe, D~A Johnston, W~Janke, and R~Kenna.
\newblock The grand-canonical asymmetric exclusion process and the one-transit
  walk.
\newblock {\em J. Stat. Mech.: Theor. Exp.}, P06001, 2004.

\bibitem{BJJK04b}
R~A Blythe, W~Janke, D~A Johnston, and R~Kenna.
\newblock {Dyck} paths, {Motzkin} paths and traffic jams.
\newblock {\em J. Stat. Mech.: Theor. Exp}, P10007, 2004.

\bibitem{CC05}
T~Castellani and A~Cavagna.
\newblock Spin-glass theory for pedestrians.
\newblock {\em J. Stat. Mech.: Theor. Exp.}, P05012, 2005.

\bibitem{Kurchan}
J~Kurchan.
\newblock Nonequilibrium work relations.
\newblock In \cite{AGT05}.

\bibitem{ES02}
D~J Evans and D~J Searles.
\newblock The fluctuation theorem.
\newblock {\em Adv. Phys.}, 51:1529, 2002.

\bibitem{Crooks00}
G~E Crooks.
\newblock Path-ensemble averages in systems driven far from equilibrium.
\newblock {\em Phys. Rev. Lett.}, 61:2361, 2000.

\bibitem{LS99}
J~L Lebowitz and H~Spohn.
\newblock A {Gallavotti-Cohen}-type symmetry in the large deviation functional
  for stochastic dynamics.
\newblock {\em J. Stat. Phys.}, 95:333, 1999.

\end{thebibliography}
\end{document}